\documentclass{bmvc2k}

\usepackage{amsmath}
\usepackage{dsfont}
\usepackage{xcolor,colortbl}
\usepackage{booktabs}  
\usepackage{multirow}  
\usepackage{multicol}  
\usepackage{siunitx}

\title{Unsupervised Feature Orthogonalization for Learning Distortion-Invariant Representations}

\addauthor{Sebastian Doerrich}{sebastian.doerrich@uni-bamberg.de}{1}
\addauthor{Francesco Di Salvo}{francesco.di-salvo@uni-bamberg.de}{1}
\addauthor{Christian Ledig}{christian.ledig@uni-bamberg.de}{1}

\addinstitution{
 xAILab Bamberg \\ 
 University of Bamberg \\
 Bamberg, Germany
}

\runninghead{Doerrich et al.}{Learning Distortion-Invariant Representations}

\definecolor{grey}{HTML}{f2f2f2}
\definecolor{lightOrange}{HTML}{F7CF9C}
\definecolor{lightPurple}{HTML}{DCBDE7}

\begin{document}

\maketitle

\begin{abstract}
This study introduces \mbox{unORANIC+}, a novel method that integrates unsupervised feature orthogonalization with the ability of a Vision Transformer to capture both local and global relationships for improved robustness and generalizability. The streamlined architecture of \mbox{unORANIC+} effectively separates anatomical and image-specific attributes, resulting in robust and unbiased latent representations that allow the model to demonstrate excellent performance across various medical image analysis tasks and diverse datasets. Extensive experimentation demonstrates \mbox{unORANIC+'s} reconstruction proficiency, corruption resilience, as well as capability to revise existing image distortions. Additionally, the model exhibits notable aptitude in downstream tasks such as disease classification and corruption detection. We confirm its adaptability to diverse datasets of varying image sources and sample sizes which positions the method as a promising algorithm for advanced medical image analysis, particularly in resource-constrained environments lacking large, tailored datasets. The source code is available at \href{https://github.com/sdoerrich97/unoranic-plus}{github.com/sdoerrich97/unoranic-plus}.
\end{abstract}

\section{Introduction}
\label{sec:intro}
\begin{figure}
    \hfil
    \subfigure{%
        \begin{minipage}{0.26\linewidth}
            \centering
            \centerline{\includegraphics[width=\linewidth]{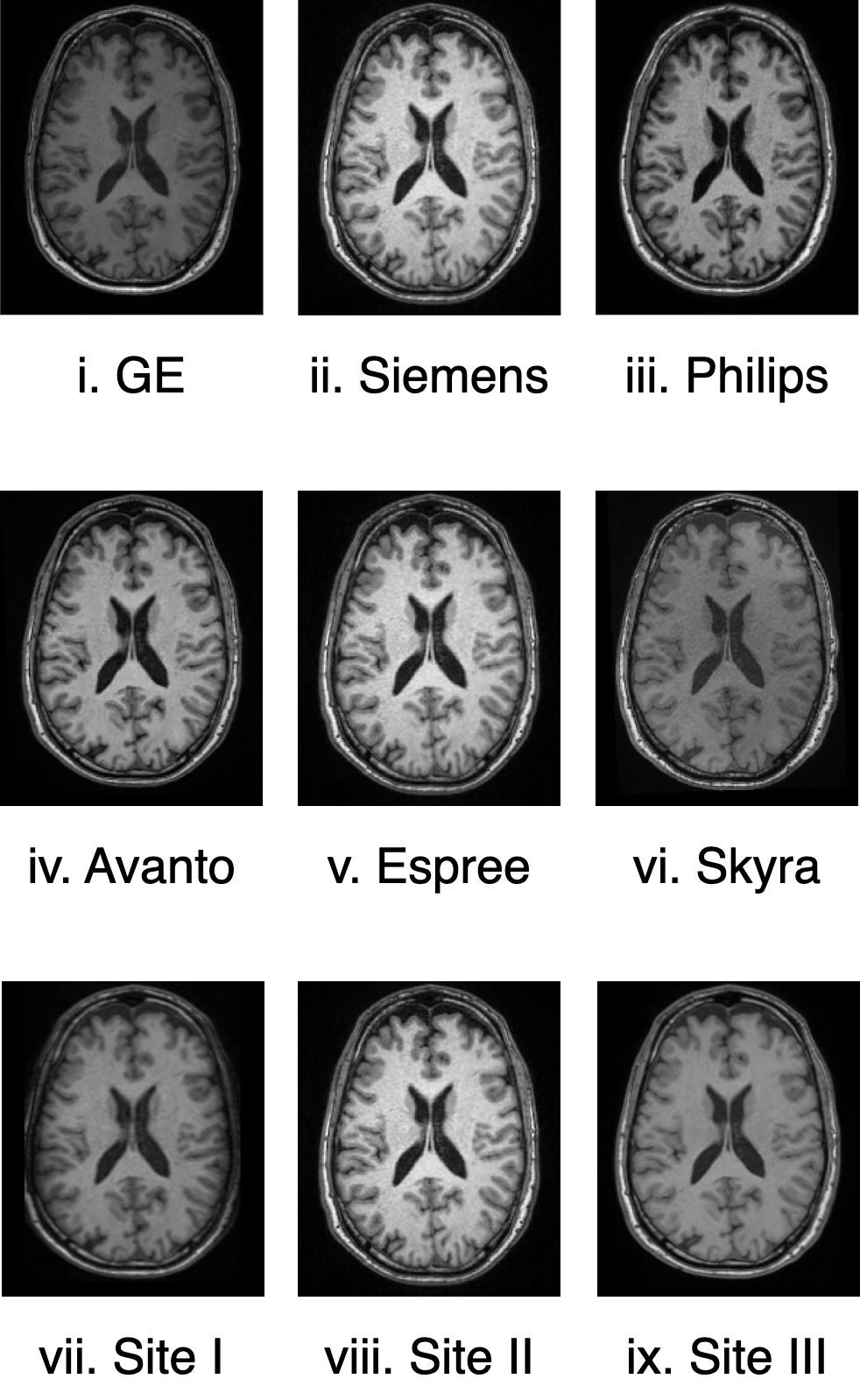}}
            \centerline{(a)}
        \end{minipage}
        \label{fig:motivation}
    }\hfil \hfil \hfil \hfil 
    \subfigure{%
        \begin{minipage}{0.65\linewidth}
            \centering
            \centerline{\includegraphics[width=\linewidth]{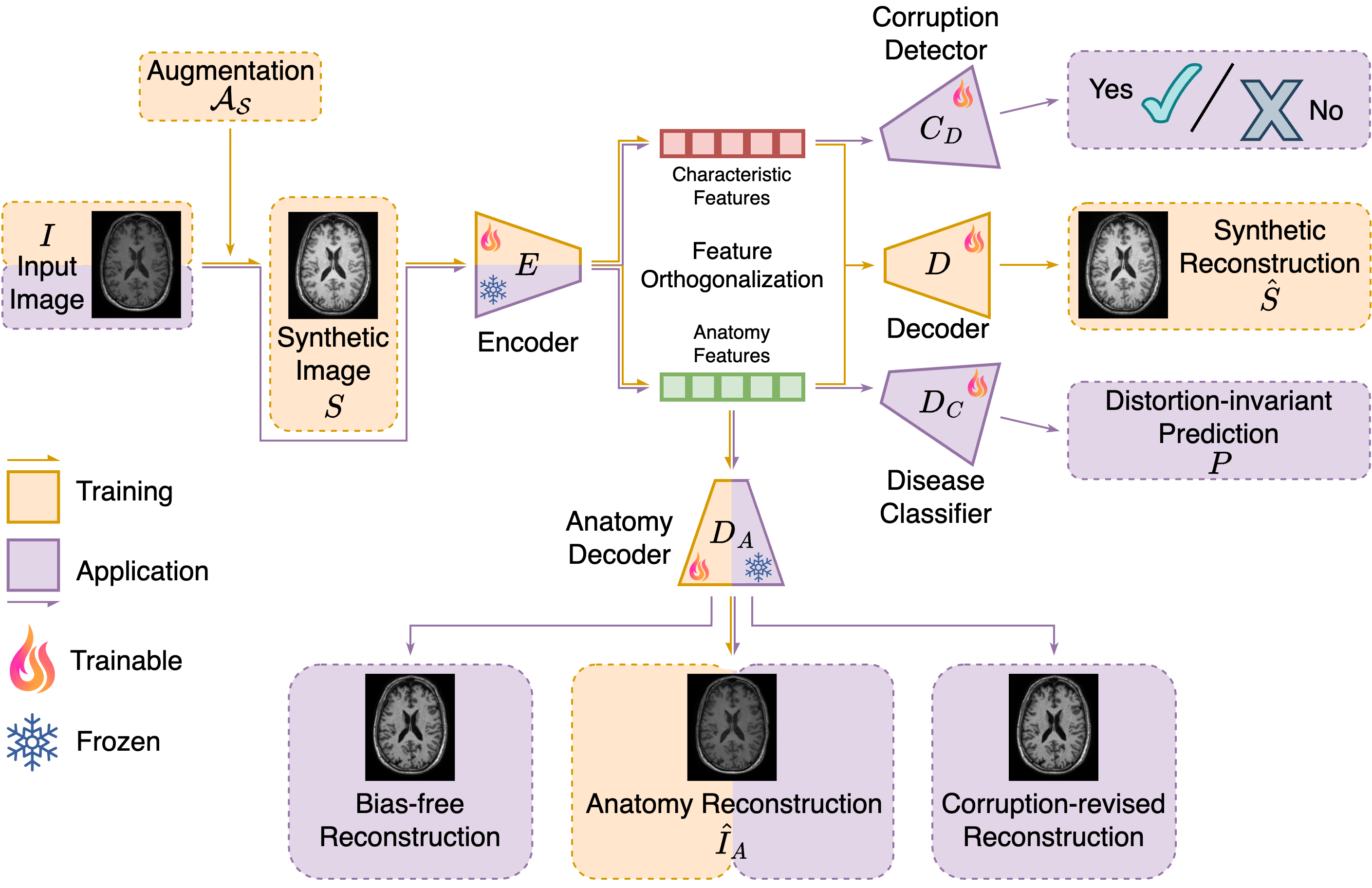}}
            \centerline{(b)}
        \end{minipage}
        \label{fig:idea}
    }
    \hfil
    \caption{(a) Illustration of domain shifts in terms of different contrasts and brightness levels among manufacturers (i-iii), among models from the same producer (iv-vi), and among the same model at different sites (vii-ix) for the same slice of the same individual across multiple scans. (b) High-level overview of our proposed approach. During training, the encoder $E$ is trained to orthogonalize anatomical and image-characteristic features in an input image ({\color{lightOrange}orange path}). Once trained, the learned feature orthogonalization by the frozen encoder is used for various downstream tasks, including bias removal, corruption detection and revision in input images, as well as robust, distortion-invariant disease classification ({\color{lightPurple}purple path}).}
    \label{fig:motviation_and_idea}
\end{figure}

In recent years, significant progress has been made in deep learning via the introduction of novel training schemes and sophisticated network architectures~\cite{Ho2020,Vaswani2017,Devlin2019}. However, achieving generalizability across diverse domains remains a challenge, limiting the impact of those advancements~\cite{Wang2023,Eche2021}. This challenge is particularly pronounced in the medical domain, where data scarcity, inhomogeneities, and underrepresented demographics hinder model effectiveness~\cite{Moor2023,Norori2021}. Moreover, domain shifts caused by variations in scanner models and imaging parameters further impede generalizability~\cite{Khan2022,Lafarge2017,Stacke2021}. An example of this can be seen in \figurename~\ref{fig:motivation}, which demonstrates distinct contrast and brightness variations across machines from different manufacturers (i–iii), different models of the same manufacturer (iv–vi), and the same model at different sites (vii–ix), despite displaying the same corresponding slice of T1-weighted MRI scans of the same healthy male subject from~\cite{Opfer2022}.

To tackle these issues, the work of \mbox{unORANIC}~\cite{Doerrich2024} has shown that unsupervised orthogonalization of anatomy and image-characteristic features can substantially improve robustness and generalizability without the need for domain knowledge, paired data, or labels. Building on this, we introduce \mbox{unORANIC+}, a simpler, more robust, and overall higher performant improvement. \figurename~\ref{fig:idea} provides a high-level overview of our proposed approach. A single encoder is trained with a reconstruction objective to orthogonalize anatomical and image-characteristic information within an input image without requiring labels, image pairs, or additional information about the image domain (orange path). Once trained, the encoder is frozen, allowing the learned feature orthogonalization to be utilized for various downstream tasks, such as unbiased anatomical image reconstruction, detection and revision of corruptions in input images, as well as robust disease classification of biased or distorted images (purple path). Consequently, this disentanglement of image information enables \mbox{unORANIC+} to generate robust latent representations even in the presence of data inhomogeneities and domain shifts. In summary, our main contributions include:

\begin{enumerate}
    \item \textbf{Enhanced feature orthogonalization:} \mbox{unORANIC+} synergizes unsupervised feature orthogonalization with a Vision Transformer's ability to capture global-local relationships for improved robustness and generalizability.
    \item \textbf{Streamlined architecture:} with a single encoder, \mbox{unORANIC+} effectively disentangles anatomical and image attributes, yielding robust latent representations to allow superior performance in a wide range of tasks.
    \item \textbf{Versatility across datasets:} extensive quantitative experimentation across various datasets and medical conditions displays \mbox{unORANIC+'s} performance and versatility.
\end{enumerate}
%
%
\section{Related work}
\label{sec:relatedwork}
\subsection{Orthogonalization of anatomy and image characteristics}
\label{subsec:orthogonalization}
Orthogonalization involves separating anatomical information from image-specific attributes in an image~\cite{Doerrich2024}. Anatomy features encompass the underlying anatomical structures like organs, tissues, or disease characteristics, while image-specific features include attributes such as contrast and brightness. This disentanglement allows for a focused assessment of anatomical elements, independent of image-specific biases or corruptions~\cite{Chartsias2019,Dewey2020ADL,Zuo2021}. The concept of an entirely unsupervised separation of these feature classes was first introduced with \mbox{unORANIC}~\cite{Doerrich2024}. \mbox{unORANIC+} builds upon this foundation, employing a Vision Transformer autoencoder to further advance that process and extend its applicability.
\subsection{Vision Transformer autoencoder}
\label{subsec:autoencodingViT}
Vision Transformers (ViTs) and autoencoding represent two key methodologies in the realm of representation learning and image processing~\cite{dosovitskiy2021,ZAKERI2024451}. ViTs excel at capturing intricate local and global relationships~\cite{Maurício2023}, offering a promising avenue for tasks like image classification, segmentation, and object detection~\cite{dosovitskiy2021,oquab2023}. Meanwhile, autoencoding is a classical technique known for its prowess in representation learning. It involves an encoder mapping inputs to a latent representation and a decoder reconstructing the original input. This process is typically conducted in a self- or unsupervised manner and holds significance in various applications including image denoising, and pre-training tasks~\cite{ZAKERI2024451,Silva2020,He2022}.

Integrating both methods, particularly by employing ViTs for the encoder and decoder of the autoencoder, harnesses the strength of ViTs in discerning complex relationships to establish a representative latent space. Furthermore, this fusion enables the training of these traditionally data-hungry ViTs in a self- or even unsupervised fashion. This approach supports learning high-capacity models with strong generalization~\cite{He2022} and enables model training for datasets with limited labels. In contrast to the asymmetric design in~\cite{He2022}, \mbox{unORANIC+} adopts a symmetric configuration to enable high-resolution reconstructions. Moreover, a second decoder and an adapted loss function are introduced to allow the reconstruction of the original input and an anatomical, bias-free version of it.
%
%
\section{Method}
\label{sec:method}
\subsection{Baseline}
\label{subsec:baseline}
\begin{figure}
    \centering
    \includegraphics[width=0.5\linewidth]{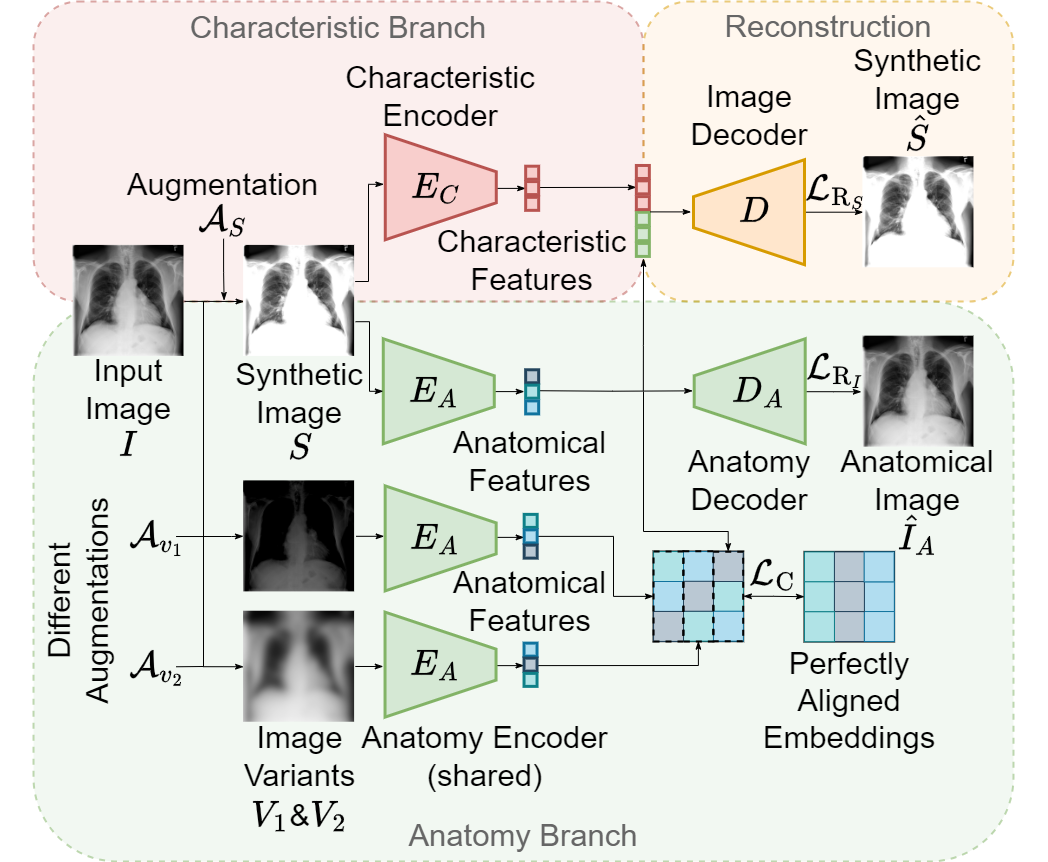}
    \caption{Schematic representation of the training pipeline for \mbox{unORANIC} (adapted from~\cite{Doerrich2024}). The input image $I$ is assumed to be bias-free and uncorrupted. Random augmentations \(\mathcal{A}_S\), \(\mathcal{A}_{v_1}\), and \(\mathcal{A}_{v_2}\) distort \( I \) to generate synthetic corrupted versions \( S \), \( V_1 \), and \( V_2 \) with identical anatomical information but different distortions. These distorted images are processed by the shared anatomy encoder \( E_A \), which uses the consistency loss \(\mathcal{L}_C\) to learn anatomical, distortion-invariant features. Concurrently, \( S \) is processed by the characteristic encoder \( E_C \) to capture image-specific details such as contrast and brightness. Reconstruction losses \(\mathcal{L}_{R_S}\) and \(\mathcal{L}_{R_I}\) are applied to the reconstructed images \(\hat{S}\) and \(\hat{I}_A\) by decoder \( D \) and \( D_A \), respectively, to ensure that $E_A$ and $E_C$ learn comprehensive, reliable features.}
    \label{fig:unoranic}
\end{figure}

The objective is to disentangle anatomy from image-specific features in an input image to learn robust, unbiased features in the latent space. For this, we harness the concept of \mbox{unORANIC} which originally employs a two-branch autoencoder~\cite{Doerrich2024}. During inference, the anatomical branch extracts core anatomical features to reconstruct a bias-free version of the input image, while the characteristic branch captures distinctive image details omitted by the first branch to allow the reconstruction of the original input image. For this, the anatomy encoder $E_A$ is shared across a set of distorted variants ($S$, $V_1$, $V_2$) of the same input image $I$ during training. $E_A$ is updated via a combination of the L2-consistency loss $\mathcal{L}_{\text{C}}$, forcing the feature embeddings of $S$, $V_1$, and $V_2$ to be the same, as well as the L2-reconstruction losses $\mathcal{L}_{\text{R}_{S}}$ and $\mathcal{L}_{\text{R}_{I}}$ to learn bias-robust anatomical representations, while the characteristic encoder $E_C$ retains image-specific details. This process is displayed in \figurename~\ref{fig:unoranic}.
\subsection{unORANIC+}
\label{subsec:unoranicplus}
In contrast to the former architecturally enforced orthogonalization, \mbox{unORANIC+} employs only one encoder $E$ which maps the input to a single, higher-dimensional latent space. The network comprises two consecutive decoders for which the first decoder $D$ reconstructs the original input image, while the other $D_A$ focuses on generating the bias-free anatomical reconstruction, akin to \mbox{unORANIC}. 
Inspired by the Masked Autoencoder (MAE) design from~\cite{He2022}, we employ Vision Transformers (ViTs) for the encoder and both decoders. Specifically, we opt for a symmetric design in which the encoder and decoders are equal in size and depth. In line with standard ViT practices~\cite{dosovitskiy2021}, we divide the image into regular non-overlapping patches. The encoder uses a linear projection with positional embeddings to embed the patches before processing them through a sequence of Transformer blocks. The two decoders will process the encoded patches through their own Transformer blocks to reconstruct the image.
\subsection{Training and application}
\label{subsec:training}
\begin{figure}
    \centering
    \includegraphics[width=0.8\linewidth]{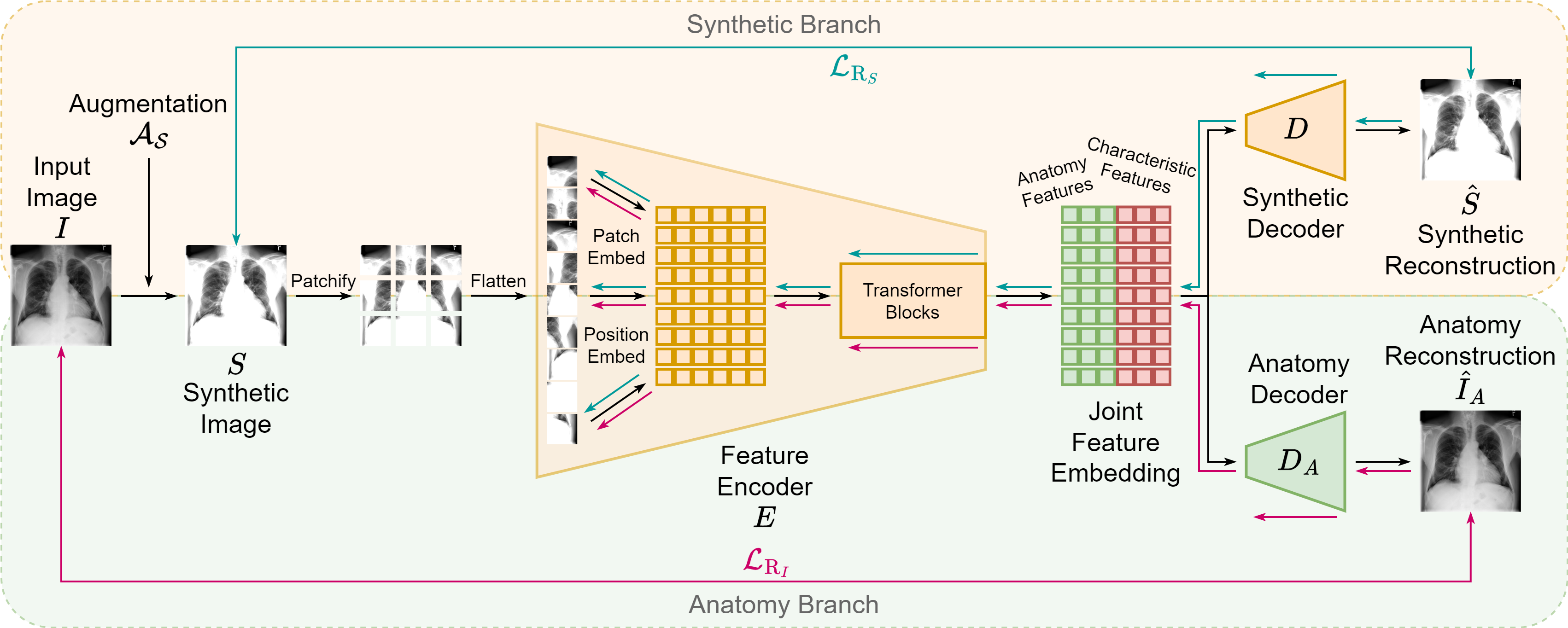}
\caption{Schematic representation of the training pipeline for the refined \mbox{unORANIC+} method on the example of chest X-ray images. The polar arrows illustrate the forward propagation and gradient flow, respectively. During training, an input image $I$ is augmented with a random set of distortions, $\mathcal{A}_S$, to generate the synthetic, distorted image $S$. $S$ is subsequently divided into non-overlapping patches before it is fed through the single Vision Transformer (ViT) encoder $E$ to map the input image to a higher-dimensional latent space. Two ViT decoders, $D$ and $D_A$, are used to reconstruct the original synthetic image $\hat{S}$ as well as a bias-free anatomical reconstruction $\hat{I}_A$, respectively. The two reconstruction losses $\mathcal{L}_{\text{R}_{S}}$ and $\mathcal{L}_{\text{R}_{I}}$ guide the separation of anatomical and image-characteristic features in the latent space and ensure a high quality of the reconstructions.}
\label{fig:architecture}
\end{figure}

Following \figurename~\ref{fig:architecture}, during training, we augment each input image $I$ with a random set of distortions $\mathcal{A_S}$, taken from~\cite{Buslaev2020}, to create a synthetic corrupted version $S$. $S$ is afterward split into regular non-overlapping patches, which are flattened into 1D vectors before passing them through the encoder $E$. Positional embeddings are added to aid the learning of spatial dependencies. The resulting embeddings are passed through the encoder’s Transformer blocks to create the latent representation, encompassing both anatomy and image-characteristic features.
This enables the synthetic decoder $D$ to reconstruct $\hat{S}$ with all added characteristics by $\mathcal{A_S}$, while the anatomy decoder $D_A$ can disregard these characteristics and reconstruct a bias-free anatomical image $\hat{I}_A$. To do so, the training exploits the two reconstruction losses $\mathcal{L}_{\text{R}_{S}}$ and $\mathcal{L}_{\text{R}_{I}}$ equally, which measure the mean squared error (MSE) in the pixel space between $\hat{S}$ and $S$ as well as $\hat{I}_A$ and $I$, respectively. The latter enforces decoder $D_A$ to focus solely on anatomical features, further promoting the feature orthogonalization within the latent space.

During inference, the model can be applied to any potentially (un-)biased or (un-)corrupted test image $I$ to orthogonalize its anatomical and image-specific features, enabling tasks like bias-free reconstruction, corruption detection and revision, or robust downstream applications such as disease classification.

%
%
%
\section{Experiments and results}
\label{sec:experiments}
\begin{figure*}
\centering
\includegraphics[width=\linewidth]{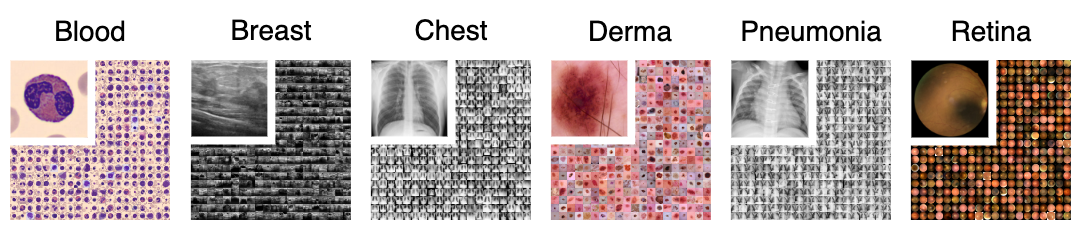}
    \caption{Examples from the datasets of the MedMNIST v2 benchmark~\cite{Yang2023} used for evaluating our approach (left to right: blood, breast, chest, derma, pneumonia, and retina)}
    \label{fig:dataset}
\end{figure*}

We comprehensively evaluate \mbox{unORANIC+} in terms of reconstruction quality, capability to revise existing corruptions, corruption robustness, and its effectiveness in downstream tasks such as disease classification and corruption detection. To allow a fair comparison with \mbox{unORANIC}, we utilize the same diverse selection of $28 \times 28$ biomedical 2D datasets from the MedMNIST v2 benchmark~\cite{Yang2023} the original method was evaluated on, including breastMNIST ($546$ training samples), retinaMNIST ($1,080$), pneumoniaMNIST ($4,078$), dermaMNIST ($7,007$), and bloodMNIST ($11,959$). Additionally, we assess all models on the larger chestMNIST dataset ($78,468$ training samples) as well. Examples for each dataset can be seen in \figurename~\ref{fig:dataset}. Finally, in addressing a major limitation of \mbox{unORANIC}, which was exclusively evaluated on $28 \times 28$ images, we investigate \mbox{unORANIC+'s} potential to handle higher dimensional data as well. This is achieved by adopting higher resolution versions of the MedMNIST datasets, comprising images of $224 \times 224$ pixels, by using the original data samples~\cite{Acevedo2020,Al-Dhabyani2020,Wang2017ChestXRay8HC,Tschandl2018,Kermany2018,Liu2022} in combination with the MedMNIST data splits.
\subsection{Reconstruction and corruption revision}
\label{sec:reconstruction}
We first evaluate the reconstruction abilities of \mbox{unORANIC+} and the original \mbox{unORANIC} model. For this, we train \mbox{unORANIC+} including its encoder and both decoders, each composed of $12$ layers and $16$ attention heads, for $150$ epochs with a batch size of $64$, using the Adam optimizer with weight decay and learning rate warmup. The patch size is set to $4 \times 4$ due to the input size of $28 \times 28$, with a latent dimension of $128$ per patch.
The results displayed in \tablename~\ref{tab:reconstructions} show the average Peak Signal-to-Noise Ratio (PSNR) and Structural Similarity Index Metric (SSIM) values on each dataset's test set for both models' anatomical reconstructions ($\hat{I}_A$) and reconstructions of the original input ($\hat{I}$) given an uncorrupted input image $I$. These findings clearly indicate that \mbox{unORANIC+} substantially outperforms the original model for both reconstruction objectives.
\begin{table}[!b]
\setlength{\tabcolsep}{4.5pt}
\begin{center}
    \begin{tabular}{l c c c c c c c c c c c c}
        \toprule
        \multirow{4}{*}{Dataset} & \phantom{x} & \multicolumn{5}{c}{PSNR} & \phantom{x} & \multicolumn{5}{c}{SSIM} \\
        \cmidrule{3-7} \cmidrule{9-13}
        & & \multicolumn{2}{c}{unORANIC} & \phantom{i} & \multicolumn{2}{c}{unORANIC+} & & \multicolumn{2}{c}{unORANIC} & \phantom{i} & \multicolumn{2}{c}{unORANIC+ }\\
        \cmidrule{3-4} \cmidrule{6-7} \cmidrule{9-10} \cmidrule{12-13}  
        & & $\hat{I}_A$ & $\hat{I}$ & & $\hat{I}_A$ & $\hat{I}$ & & $\hat{I}_A$ & $\hat{I}$ & & $\hat{I}_A$ & $\hat{I}$ \\
        \midrule
        Blood & & $27.06$ & $31.70$ & & $\mathbf{35.88}$ & $\mathbf{49.15}$ & & 0.877 & 0.943 & & \textbf{0.987} & \textbf{0.999} \\
        Breast & & $19.48$ & $29.39$ & & $\mathbf{26.21}$ & $\mathbf{33.55}$ & & 0.526 & 0.816 & & \textbf{0.889} & \textbf{0.957} \\
        Chest & & $27.93$ & $33.73$ & & $\mathbf{35.30}$ & $\mathbf{56.02}$ & & 0.956 & 0.983 & & \textbf{0.995} & \textbf{0.999} \\
        Derma & & $23.73$ & $38.57$ & & $\mathbf{30.07}$ & $\mathbf{45.09}$ & & 0.864 & 0.970 & & \textbf{0.971} & \textbf{0.995} \\
        Pneumonia & & $24.00$ & $36.04$ & & $\mathbf{28.96}$ & $\mathbf{44.80}$ & & 0.901 & 0.977 & & \textbf{0.977} & \textbf{0.997} \\
        Retina & & $27.50$ & $36.31$ & & $\mathbf{30.39}$ & $\mathbf{37.71}$ & & 0.888 & 0.954 & & \textbf{0.936} & \textbf{0.978} \\
        \bottomrule
    \end{tabular}
\end{center}
\caption{Comparison of average Peak Signal-to-Noise Ratio (PSNR) and Structural Similarity Index Metric (SSIM) values for the anatomical reconstructions ($\hat{I}_A$) and the reconstructions of the original input ($\hat{I}$) given an uncorrupted input image $(I)$ between \mbox{unORANIC} and \mbox{unORANIC+}. The best performance per reconstruction task is indicated in \textbf{bold}.}
\label{tab:reconstructions}
\end{table}

\begin{figure}
    \hfil
    \subfigure{%
        \begin{minipage}[b]{0.57\linewidth}
           \centering
           \centerline{\includegraphics[width=\linewidth]{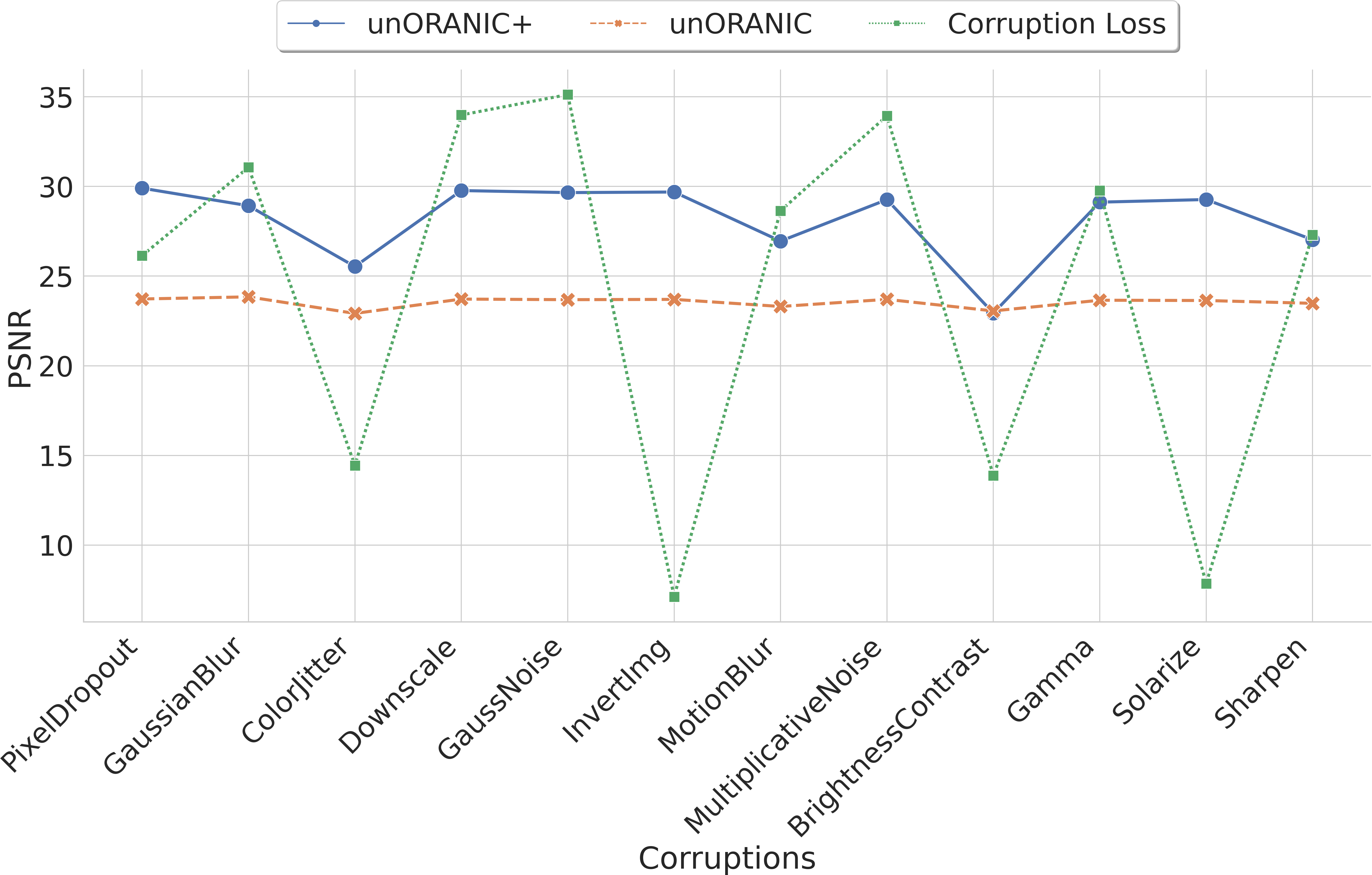}}
           \centerline{(a)}
        \end{minipage}
        \label{fig:corruption_wise_revision}
    }
    \hfil
    \subfigure{%
        \begin{minipage}[b]{0.215\linewidth}
            \centering
            \centerline{\includegraphics[width=\linewidth]{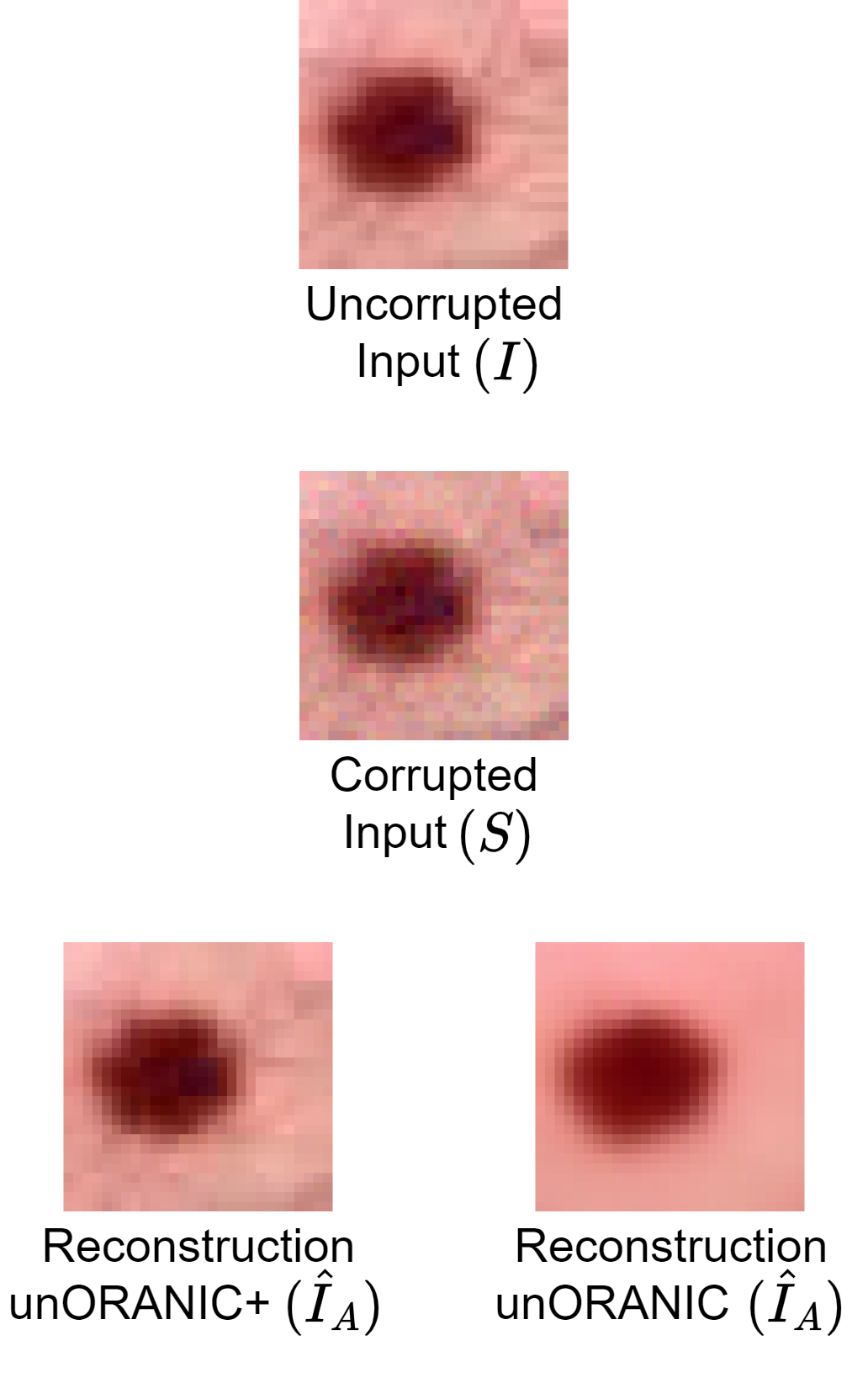}}
            \centerline{(b)}
        \end{minipage}
        \label{fig:strong_corruption_revision}
    }
    \hfil
    \caption{Corruption revision capabilities of \mbox{unORANIC} and \mbox{unORANIC+}. In (a), their reconstruction consistency is depicted despite the corruption-related image quality loss (PSNR between the original image $I$ and the distorted variant $S$ - "green dotted line"). (b) highlights the distortion correction capabilities of both methods using Gaussian noise as an example.}
    \label{fig:corruptions_revision}
\end{figure}

Additionally, we assess both models based on their capability to revise corruptions in an input image via their anatomy branches. For this, we deliberately apply the same range of corruptions as used during training to all test images to generate synthetic distorted versions, similar to the experiment presented in~\cite{Doerrich2024}. Despite the distortions, both models successfully reconstruct the uncorrupted input images, as depicted in \figurename~\ref{fig:corruptions_revision} for the dermaMNIST dataset. Moreover, the figure illustrates \mbox{unORANIC+'s} enhanced corruption revision capability across all distortions in \ref{fig:corruption_wise_revision} and emphasizes its ability to preserve fine-grained details in \ref{fig:strong_corruption_revision} compared to its predecessor for distortions such as the displayed Gaussian noise.
\subsection{Disease classification and corruption detection}
\label{sec:classification}
We evaluate the encoded feature embeddings of \mbox{unORANIC+} to assess the degree of orthogonalization between anatomy and image-characteristic features for the tasks of disease classification and corruption detection. Depending on the used dataset, the disease classification comprises a binary classification (breast, chest, pneumonia) or a multi-class classification (blood, derma, retina) task. To this end, we freeze the encoder, replace the decoders with an architecturally alike ViT classifier for the particular task and dataset, and train only the classifier. The model's performance, measured in terms of Accuracy (ACC) and Area Under the ROC Curve (AUC) per dataset, is compared with the \mbox{unORANIC} model and supervised, end-to-end trained ResNet-18 and ViT baselines for each respective task. It is important to note that while these baselines provide a useful context for performance assessment, their results should not be compared directly to those of \mbox{unORANIC} and \mbox{unORANIC+}. Both baselines are trained fully supervised, end-to-end, while \mbox{unORANIC} and \mbox{unORANIC+} both employ frozen encoder(s) and only train the classifier head for the specific task. The results in \tablename~\ref{tab:classification} for bloodMNIST demonstrate unORANIC+'s notable improvement over unORANIC, along with its comparable performance to the end-to-end trained, supervised models.

\begin{table}[h]
\begin{center}
    \begin{tabular}{l c c c c c c c c c c c c}
        \toprule
        \multirow{2.5}{*}{Methods} & \phantom{i} & \multicolumn{5}{c}{Disease Classification} & \phantom{xx} & \multicolumn{5}{c}{Corruption Detection} \\
        \cmidrule{3-7} \cmidrule{9-13}
        & & & ACC & & AUC & & & & ACC & & AUC & \\
        \midrule
        \cellcolor{grey} $\text{ResNet-18}^\dag$ & \cellcolor{grey} & \cellcolor{grey} & \cellcolor{grey} 0.958 & \cellcolor{grey} & \cellcolor{grey} 0.998 & \cellcolor{grey} & \cellcolor{grey} & \cellcolor{grey} & \cellcolor{grey} 0.973 & \cellcolor{grey} & \cellcolor{grey} 0.998 & \cellcolor{grey} \\
        \cellcolor{grey} $\text{ViT}^\dag$ & \cellcolor{grey} & \cellcolor{grey} & \cellcolor{grey} 0.930 & \cellcolor{grey} & \cellcolor{grey} 0.992 & \cellcolor{grey} & \cellcolor{grey} & \cellcolor{grey} & \cellcolor{grey} 0.914 & \cellcolor{grey} & \cellcolor{grey} 0.926 & \cellcolor{grey} \\
        unORANIC          & & & 0.800 & & 0.952 & & & & 0.962 & & 0.976 & \\
        unORANIC+ & & & \textbf{0.935} & & \textbf{0.994} & & & & \textbf{0.970} & & \textbf{0.980} & \\
        \bottomrule
    \end{tabular}
\end{center}
\caption{Comparison of the classification and corruption detection results on the bloodMNIST dataset. Fully supervised models, trained end-to-end, are indicated with $\vphantom{x}^\dag$. A superior performance of \mbox{unORANIC+} compared to \mbox{unORANIC} is indicated in \textbf{bold}.}
\label{tab:classification}
\end{table}

\begin{figure*}
\centering
\includegraphics[width=0.78\linewidth]{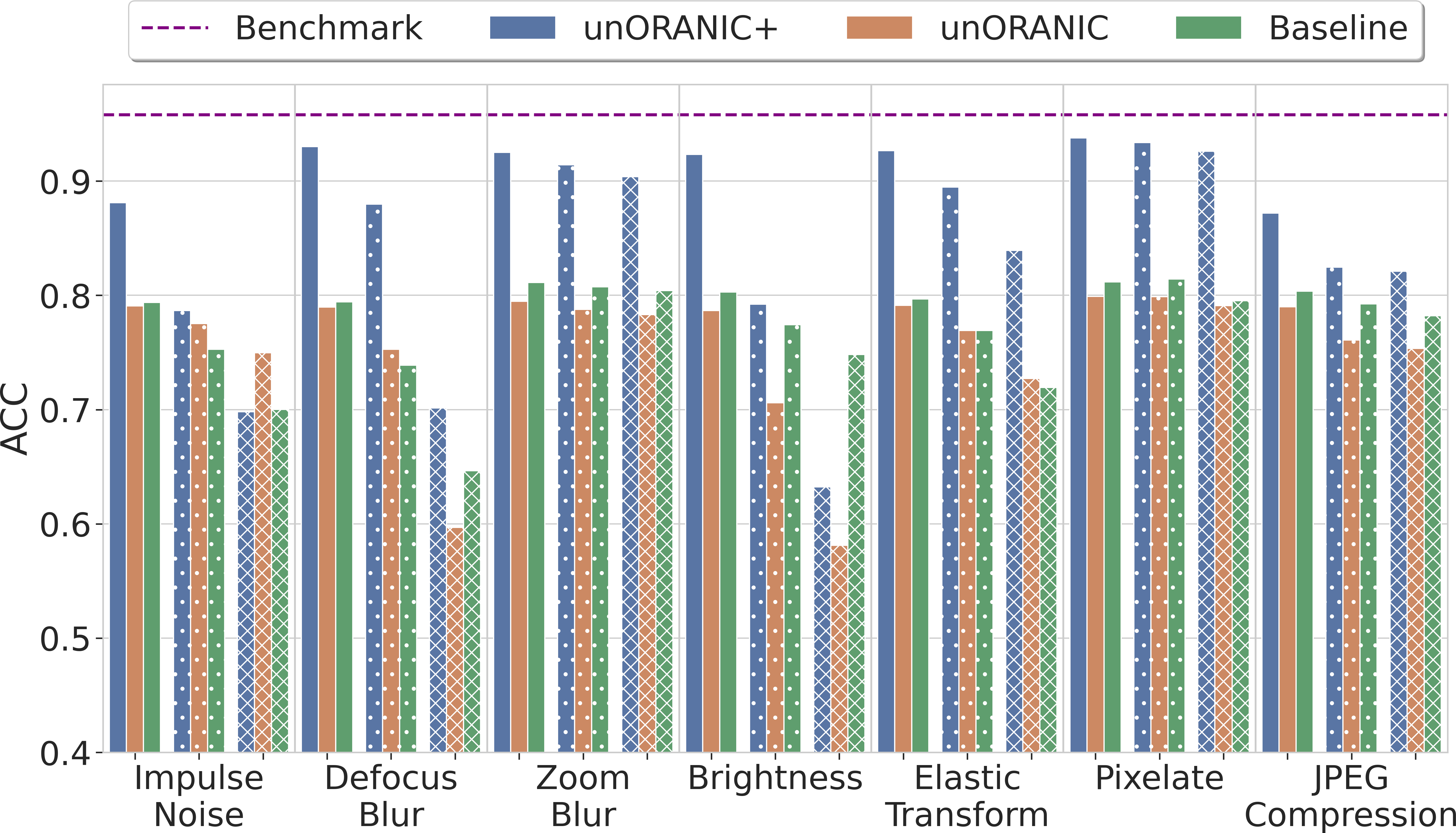}
    \caption{Visualization of \mbox{unORANIC+'s} resilience to unseen corruptions compared to the reference models (unORANIC and the supervised ResNet-18 "Baseline"), demonstrated by its robust disease classification performance on the bloodMNIST dataset as corruption severity increases. Different textures (\phantom{ $\cdot$}, $\cdot$ , $\times$) indicate distinct levels of severity for each presented corruption. For reference, the plot also includes the supervised classification "Benchmark" (i.e., ResNet-18) for uncorrupted images.}
    \label{fig:robustness}
\end{figure*}

Furthermore, we test \mbox{unORANIC+'s} robustness to unseen corruptions that were not used during training, by applying varying severities of these corruptions to all test images before passing them through the trained models for the original disease classification task. We observe superior resilience of \mbox{unORANIC+} compared to \mbox{unORANIC} and ResNet-18 ("Baseline") across all datasets, as illustrated in \figurename~\ref{fig:robustness} for bloodMNIST.

\subsection{Evaluation on higher dimensional datasets}
\label{sec:realdata}
For the evaluation of \mbox{unORANIC+} on the higher dimensional versions of the MedMNIST datasets, we adjusted the ViT configuration (using $16 \times 16$ patches instead of $4 \times 4$ and a latent dimension of $768$ instead of $128$). We retrained the model for each dataset individually and conducted the same experiments as previously described. \tablename~\ref{tab:high-res-recon} presents the evaluation for the reconstruction of the original input $\hat{I}$ and the anatomical, bias-free version $\hat{I}_A$ in terms of PSNR. Additionally, \tablename~\ref{tab:high-res-class} presents the evaluation for disease classification and corruption detection (both in terms of AUC) for each individual dataset. To facilitate side-by-side comparisons across all six datasets, we omit the SSIM results for reconstruction and the ACC results for classification and detection, respectively. We remind that the $\text{ResNet-18}^\dag$ and $\text{ViT}^\dag$ reference models are trained end-to-end in a supervised manner solely for the classification and detection tasks. Therefore, we do not evaluate these models for reconstruction and only compare \mbox{unORANIC+} with its predecessor in this case.

\mbox{unORANIC+} maintains its accurate performance, showcasing robust anatomical reconstruction capabilities across all datasets and reliable corruption revision (consistently high PSNR between $\hat{I}_A$ and $\hat{I}$). Moreover, \mbox{unORANIC+} excels beyond its already robust disease classification potential for the undistorted blood and chest datasets, even exceeding the end-to-end trained baselines. This performance boost can be attributed to the larger image sizes, enabling more effective learning of both global and local representations.

While \mbox{unORANIC+} preserves a high corruption detection performance across all datasets, the disease classification results reveal limitations of ViT-based architectures for comparatively small but higher dimensional datasets, such as breast, derma, and retina. Despite \mbox{unORANIC+} surpassing the end-to-end trained ViT classifier, both lag behind the convolutional methods for the clean, undistorted versions of these datasets. Thus, it appears that the robust representations learned by \mbox{unORANIC+} are particularly beneficial for medium-sized and potentially biased or distorted datasets of small or higher image dimension.

\begin{table}
\setlength{\tabcolsep}{3.1pt}
\renewcommand{\arraystretch}{1.11}
{\scriptsize
\begin{center}
    \begin{tabular}{l cccccc c cccccc}
        \toprule

	& \multicolumn{6}{c}{Input Reconstruction $\left(\hat{I}\right)$} & & \multicolumn{6}{c}{Anatomical Reconstruction    $\left(\hat{I}_A\right)$} \\
        & \multicolumn{6}{c}{[PSNR]} & & \multicolumn{6}{c}{[PSNR]} \\
        \cmidrule(lr){2-7} \cmidrule(lr){9-14}
        & Blood & Breast & Chest & Derma & Pneumonia & Retina & & Blood & Breast & Chest & Derma & Pneumonia & Retina \\
        \midrule
        unORANIC  & 26.66 & 20.28 & 28.33 & 27.55 & 27.21 & 31.25 &   & 24.57 & 17.62 & 26.02 & 23.10 & 21.61 & 25.64 \\
        unORANIC+ & \textbf{44.63} & \textbf{27.80} & \textbf{42.23} & \textbf{33.22} & \textbf{34.87} & \textbf{35.37} &   & \textbf{38.23} & \textbf{24.94} & \textbf{32.35} & \textbf{26.91} & \textbf{26.86} & \textbf{31.41} \\
        \bottomrule
    \end{tabular}
\end{center}
}
\caption{Comparison of average Peak Signal-to-Noise Ratio (PSNR) for the reconstructions of the original input ($\hat{I}$) and the clean anatomical reconstructions ($\hat{I}_A$) given an input image ($I$) between unORANIC and unORANIC+ on the test sets of all six higher dimensional datasets. The best performance per reconstruction task is indicated in \textbf{bold}.}
\label{tab:high-res-recon}
\end{table}
\begin{table}
\setlength{\tabcolsep}{3.1pt}
\renewcommand{\arraystretch}{1.11}
{\scriptsize
\begin{center}
    \begin{tabular}{l cccccc c cccccc}
        \toprule
        & \multicolumn{6}{c}{Disease Classification} & & \multicolumn{6}{c}{Corruption Detection} \\
        & \multicolumn{6}{c}{[AUC]} & & \multicolumn{6}{c}{[AUC]} \\
        \cmidrule(lr){2-7} \cmidrule(lr){9-14}
        & Blood & Breast & Chest & Derma & Pneumonia & Retina & & Blood & Breast & Chest & Derma & Pneumonia & Retina \\

        \midrule

        \cellcolor{grey} $\text{ResNet-18}^\dag$ & \cellcolor{grey} 0.840 & \cellcolor{grey} 0.794 & \cellcolor{grey} 0.518 & \cellcolor{grey} 0.608 & \cellcolor{grey} 0.977 & \cellcolor{grey} 0.684 & \cellcolor{grey} & \cellcolor{grey} 0.941 & \cellcolor{grey} 0.833 & \cellcolor{grey} 0.876 & \cellcolor{grey} 0.842 & \cellcolor{grey} 0.867 & \cellcolor{grey} 0.869 \\
        \cellcolor{grey} $\text{ViT}^\dag$       & \cellcolor{grey} 0.891 & \cellcolor{grey} 0.580 & \cellcolor{grey} 0.535 & \cellcolor{grey} 0.501 & \cellcolor{grey} 0.840 & \cellcolor{grey} 0.647 & \cellcolor{grey} & \cellcolor{grey} 0.791 & \cellcolor{grey} 0.663 & \cellcolor{grey} 0.820 & \cellcolor{grey} 0.660 & \cellcolor{grey} 0.639 & \cellcolor{grey} 0.700 \\
        unORANIC                & 0.930 & 0.774 & 0.528 & 0.717 & 0.953 & 0.681 &   & 0.724 & 0.586 & 0.635 & 0.625 & 0.639 & 0.644 \\
        unORANIC+               & \underline{\textbf{0.997}} & 0.757 & \underline{\textbf{0.563}} & 0.563 & \textbf{0.955} & 0.675 &   & \underline{\textbf{0.954}} & \textbf{0.616} & \underline{\textbf{0.892}} & \textbf{0.690} & \textbf{0.694} & \textbf{0.783} \\

        \bottomrule
    \end{tabular}
\end{center}
}
\caption{Comparison of the disease classification and corruption detection results across the higher dimensional datasets. Fully supervised models, trained end-to-end, are indicated with $\vphantom{x}^\dag$. \textbf{Bold} highlights superior performance of \mbox{unORANIC+} compared to \mbox{unORANIC}, while \underline{underlinin}g indicates cases where \mbox{unORANIC+} even surpasses the best supervised baseline.}
\label{tab:high-res-class}
\end{table}
\section{Discussion and Conclusion}
\label{sec:conclusion}
Through the integration of Vision Transformers for unsupervised feature orthogonalization, \mbox{unORANIC+} effectively disentangles anatomical and image-characteristic features, yielding robust latent representations. Our experiments across distinct datasets from various modalities, all showing different medical conditions, demonstrate superior reconstruction and corruption revision capabilities compared to the original \mbox{unORANIC} model, displaying its stable resilience against distortions. Additionally, we demonstrate its prowess in disease classification and corruption detection tasks as well as its adaptability to higher dimensional datasets, underscoring its potential for robust medical image analysis.

\subsubsection*{Compliance with ethical standards}
This research study was conducted retrospectively using human subject data made available in open access by~\cite{Opfer2022} and~\cite{Yang2023}. Ethical approval was not required as confirmed by the license attached with the open-access data.

\subsubsection*{Acknowledgments}
This study was funded through the Hightech Agenda Bayern (HTA) of the Free State of Bavaria, Germany.

\subsubsection*{Disclosure of Interests}
The authors have no competing interests to declare that are relevant to the content of this article.

\bibliography{egbib}
\end{document}